\documentclass[a4paper,11pt]{article}
\usepackage{pos}
\usepackage{float}
\usepackage{graphicx}
\usepackage{subcaption}
\usepackage[font=small]{caption}
\usepackage{comment}

\title{Gamma/hadron discriminant variables in application to high-energy cosmic-ray air showers}
\ShortTitle{Gamma/hadron discriminant variables}

\author*[a]{Nataliia Borodai}

\affiliation[a]{Institute of Nuclear Physics, Polish Academy of Sciences,\\
  Krakow, Poland}

\emailAdd{nataliia.borodai@ifj.edu.pl}

\abstract{

Identification of primary cosmic rays on an event-by-event basis is a much-desired capability of cosmic-ray observatories. Several cosmic-ray air-shower experiments use so-called photon tags for gamma/hadron primary particle discrimination. These photon tag variables are derived from the total signals measured by an array of detectors and are correlated with the total number of muons in the air shower. In this work, variables based on time distribution of signals in detectors (trace-based discriminant variables) are studied and compared to total-signal-based variables. This study relies on simulated high-energy cosmic-ray air showers with energies around $10^{17.5} eV$. Since the variables discussed are derived from total signals and their time traces, which can be directly measured in real data, they are suitable for use as discriminant variables in the real ground-based cosmic ray experiments.
}

\ConferenceLogo{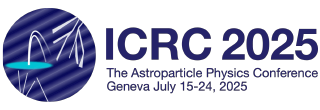}

\FullConference{39th International Cosmic Ray Conference (ICRC2025)\\
 15–24 July 2025\\
Geneva, Switzerland\\}

\begin{document}
\maketitle

\section{Introduction}

\vspace{-0.5cm} 
Event-by-event identification of primary cosmic-ray particles represents a fundamental challenge in modern astroparticle physics. One aspect of this challenge lies in distinguishing electromagnetic showers initiated by gamma rays from hadronic cascades produced by charged cosmic-ray nuclei, which differ in their atmospheric development and detector response. The ability to make this distinction is essential for extracting gamma-ray signals from the overwhelming hadronic background in ground-based observatories.

For TeV to PeV energies, gamma-hadron discriminants were commonly developed in ground-based experiments such as Tibet AS$\gamma$, LHAASO, and HAWC observatories using techniques like muon content analysis and azimuthal non-uniformity methods. The Cherenkov Telescope Array Observatory (CTAO) employs imaging atmospheric Cherenkov telescopes with gamma-hadron separation cuts and advanced discrimination methods analyzing shower signal footprint patterns, while IceCube's IceTop exploits the correlation between surface shower energy and deep ice muon deposits for cosmic ray composition analysis. It is also important to develop gamma/hadron discriminants for higher energies (above $10^{17}$ eV). The transition from PeV to EeV energies represents the ultimate frontier where cosmic rays carry energies seven orders of magnitude beyond the LHC, creating air showers of unprecedented complexity that probe the fundamental limits of particle acceleration, with discrimination becoming connected with cosmological processes like the GZK cutoff that fundamentally limits ultra-high-energy photon propagation from distant sources.

Traditional gamma/hadron discrimination relies on established variables exploiting physical differences between shower types. The atmospheric depth of shower maximum ($X_{max}$) serves as a primary photon tag discriminant \cite{ref1}, with electromagnetic cascades reaching predictable depths while hadronic showers exhibit deeper, more variable profiles. Another photon tag variable derived from integrated signal responses (total-signal-based variable) is the shower width parameter ($S_b$), which quantifies lateral distribution width \cite{ref2}, as electromagnetic showers display compact lateral profiles compared to broader hadronic distributions. Additionally, the tail probability $P_{tail}$ can be classified as a total-signal-based discriminant variable,  which measures the fraction of signal contained in the outer regions of the lateral distribution \cite{ref3}. 
These photon tag variables ($S_b$ and $P_{tail}$), derived from total signal measurements across detector arrays, correlate with the muon content of air showers. Since electromagnetic showers produce minimal muonic components while hadronic interactions generate substantial muon quantities, these total-signal-based discriminants provide robust separation capabilities. 

However, the photon tag discriminants discussed above are primarily based on geometric and integral shower properties that may not fully exploit the temporal information available in modern detector systems. So, the trace-based discriminant variable, derived from the detailed time distribution of detector signals (signal traces), is explored to offer new opportunities to enhance discrimination performance beyond conventional approaches. This analysis focuses on a detailed study of the trace-based discriminant variable, and its comparison to conventional photon tag variables. A trace-based  discriminant variable was studied first in \cite{ref4}. The analysis, presented in this paper, is different, but also based on a study of the $P_{tail}$ discriminant.

\section{Detector and Dataset}
The development of improved discriminant variables requires careful validation through Monte Carlo simulations \cite{ref5}. This work involves simulations of extensive air showers at high energies, and their reconstruction on a ground-based detector array to evaluate the performance of both conventional and trace-based discriminant variables under realistic experimental conditions.

\begin{figure}[H]
\vspace{-0.4cm} 
    \centering
    \begin{minipage}[b]{0.45\textwidth}
        \centering
        \includegraphics[width=\textwidth]{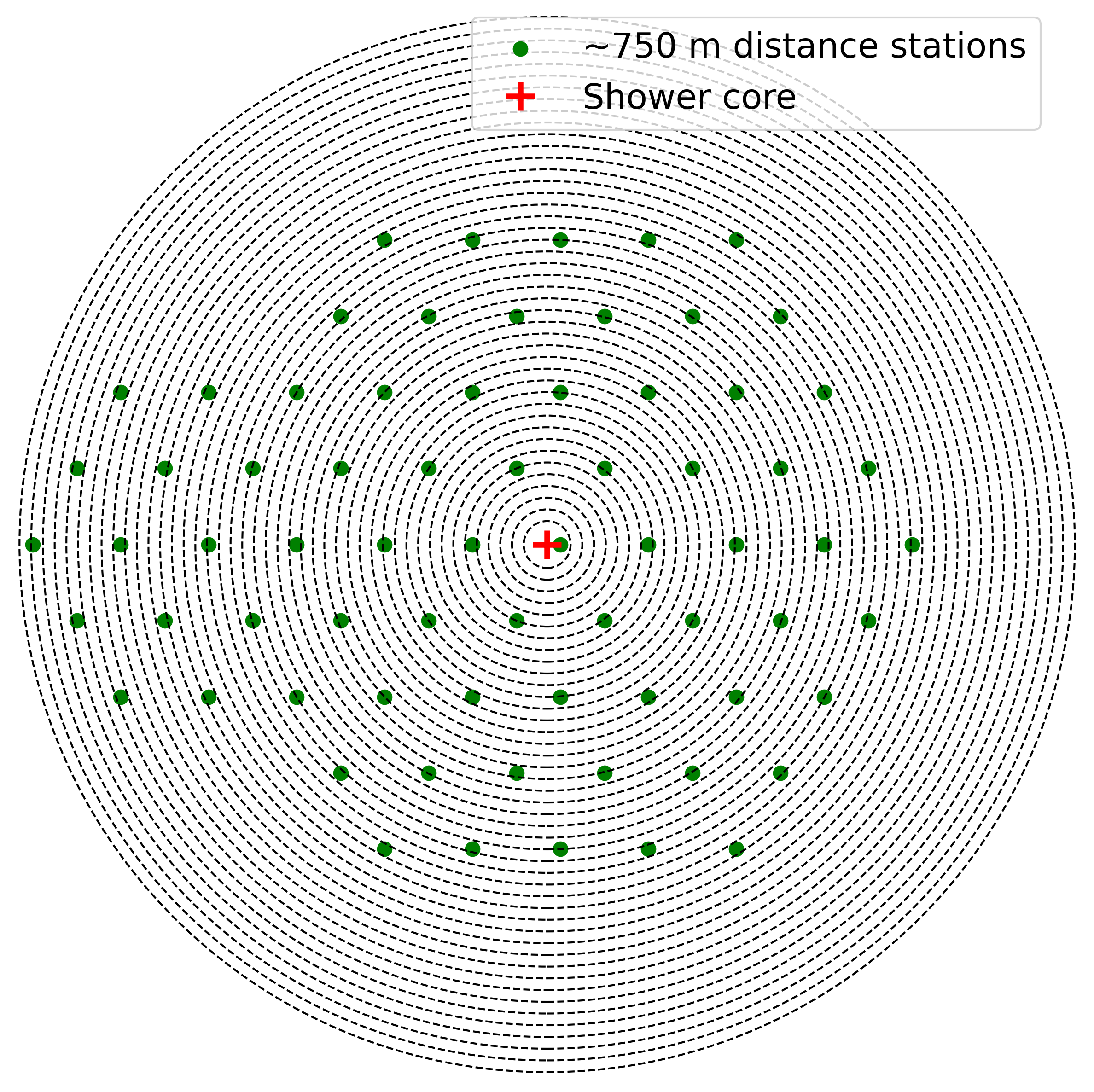}
        \caption{An exemplary array for this analysis.}
        \label{fig:1-rings}
    \end{minipage}
    \hspace{0.5cm} 
    \begin{minipage}[b]{0.45\textwidth}
        \centering
        \includegraphics[width=\textwidth]{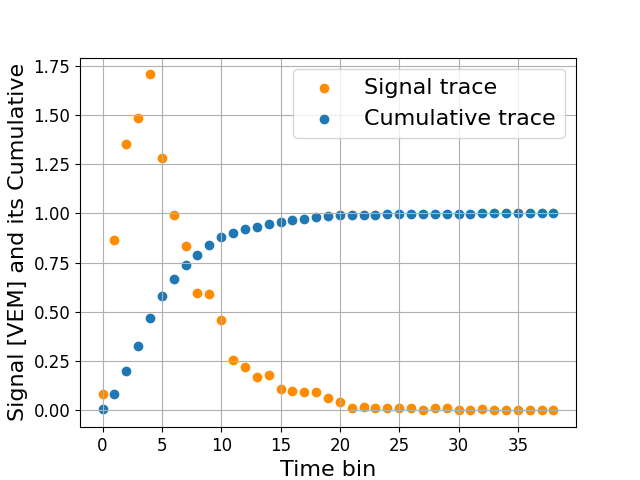}
        \caption{Signal trace $S_i(k)$ - time distribution of instantaneous signal in VEM units (Vertical Equivalent Muon) and its cumulative distribution $C_i(k)$, $k$ represents time bin numbers, $i$ is a station number.}
        \label{fig:2-trace}
    \end{minipage}
\end{figure}

The detector array used in this study  (Auger Infill array \cite{ref6}) consists of water Cherenkov detector (WCD) stations (Figure \ref{fig:1-rings}) arranged in a regular grid with 750 m spacing between the surface detectors. This configuration of the detector array enables accurate reconstruction of extensive air shower parameters and provides the necessary sensitivity for discriminant variable analysis at high energies. 

\vspace{-0.4cm} 

The air shower simulations were performed using CORSIKA version 7.8010 (COsmic Ray SImulations for KAscade) Monte Carlo package \cite{ref5} to create a robust dataset for gamma/hadron discrimination studies. The hadronic interaction modeling employed EPOS-LHC-R \cite{ref7} for modeling high-energy hadronic interactions and URQMD 1.3cr \cite{ref8} for low-energy nuclear processes, covering the complete range of nuclear physics mechanisms for atmospheric shower development. 

Over 14,000 simulations of extensive air showers were produced, with equal contributions from gamma-ray and proton primaries to ensure unbiased discriminant performance assessment. The primary cosmic-ray particles were simulated with a fixed energy of $E_{\rm{sim}} = 10^{17.5}$ eV across zenith angles ranging from 0° to 60°, matching the angular coverage of the standard reconstruction mode of conventional ground-based detectors and enabling detailed analysis of discriminant performance across the full range of shower arrival directions.

The simulated air showers were reconstructed using the 750 m spacing detector array geometry, employing standard extensive air shower analysis algorithms for core position determination, arrival direction reconstruction, and energy estimation, using the Offline Software \cite{ref9}, developed by the Pierre Auger Collaboration. Following reconstruction, approximately 12,000 events were selected based on energy criteria, This selection process focused on events with reconstructed energies within the range $E_{\rm{rec}} = [10^{17.2}$ eV, $10^{17.8}$ eV$]$, centered around the target simulation energy. 

\section{Methodology Description}
The methodology used in this study introduces $C_{tail}$ - a novel cumulative tail analysis method for gamma/hadron discrimination in extensive air showers. While conceptually related to the $P_{tail}$ method studied in \cite{ref3} and \cite{ref4}, $C_{tail}$ represents a different approach that focuses on cumulative signal distributions rather than probability-based analysis. The $C_{tail}$ method is related to the existing $P_{tail}$ method, but works differently. $P_{tail}$ looks at individual detector stations to see if they receive unusually strong signals. $C_{tail}$ instead looks at the entire shower pattern to see if the whole tail region shows unusual behavior. Both methods work because gamma rays and protons create different types of particle showers - gamma showers are more uniform and predictable, while proton showers are more irregular and variable.
The $C_{tail}$ method encompasses the following steps: spatial distribution study, trace-based (i.e. time distribution) cumulative analysis for reconstructed events, optimization, and comparison to conventional variables. $C_{tail}$ uses multi-level cumulative analysis to characterize the temporal evolution of signal traces at individual stations before aggregating them into event-level discriminants.

\textbf{Spatial distribution study.}
For analysis of spatial distribution of detector responses, the detector array surface is divided into concentric rings of equal width $100 \:m$, centered around the reconstructed shower core position (Figure \ref{fig:1-rings}). The shower core of an extensive air shower recorded by this array is located within the array and is indicated by a red cross. The ring structure begins at a minimum distance of $200 \:m$ from the shower core to avoid potential saturation effects in the detector responses. Shower cores from all events are superimposed to collect stations at equivalent distances from the shower axis. The distance between the detector and the shower axis is calculated separately for each event, taking into account its zenith angle.

\textbf{Trace-based cumulative analysis for reconstructed events}. A trace $S_i(k)$ represents the signal time distribution plotted for discrete time bins (Figure \ref{fig:2-trace}), where $S_i(k)$ is the trace signal at time $k$ for station $i$. The temporal analysis begins at the arrival time determined from the shower core reconstruction. For each detector station, traces are extracted, and the cumulative distribution of signal is calculated for each trace.
The normalized cumulative signal at station $i$ up to time bin $k$ is defined as:
\vspace{-0.4cm} 
\begin{equation}
    C_i(k) = \frac{1}{S_i^{\text{total}}} \sum_{j=1}^{k} S_i(j),
\end{equation}

\noindent where $S_i^{\text{total}} = \sum_{j=1}^{N} S_i(j)$ is the total integrated signal at station $i$ over all $N$ time bins. This normalization ensures that the cumulative distribution is independent of the absolute signal amplitude, focusing purely on the temporal structure of the shower development.

In the next step the station-level discriminator is computed by summing all normalized cumulative values across time bins in each trace. This formula can be simplified to its most elegant form: 

\vspace{-0.7cm} 
\begin{equation}
    C_i = \sum_{k=1}^{N} C_i(k) = \frac{1}{S_i^{\text{total}}} \sum_{j=1}^{N} (N - j + 1) S_i(j).
\end{equation}

The weighting function $w_j = N - j + 1$ creates a temporal emphasis where earlier bins receive higher individual weights, but the cumulative nature ensures that extended temporal profiles (characteristic of hadronic showers) produce larger discriminator values.

\begin{figure}[H]
  \begin{subfigure}[b]{0.48\linewidth}
    \centering
    \includegraphics[width=0.85\linewidth]{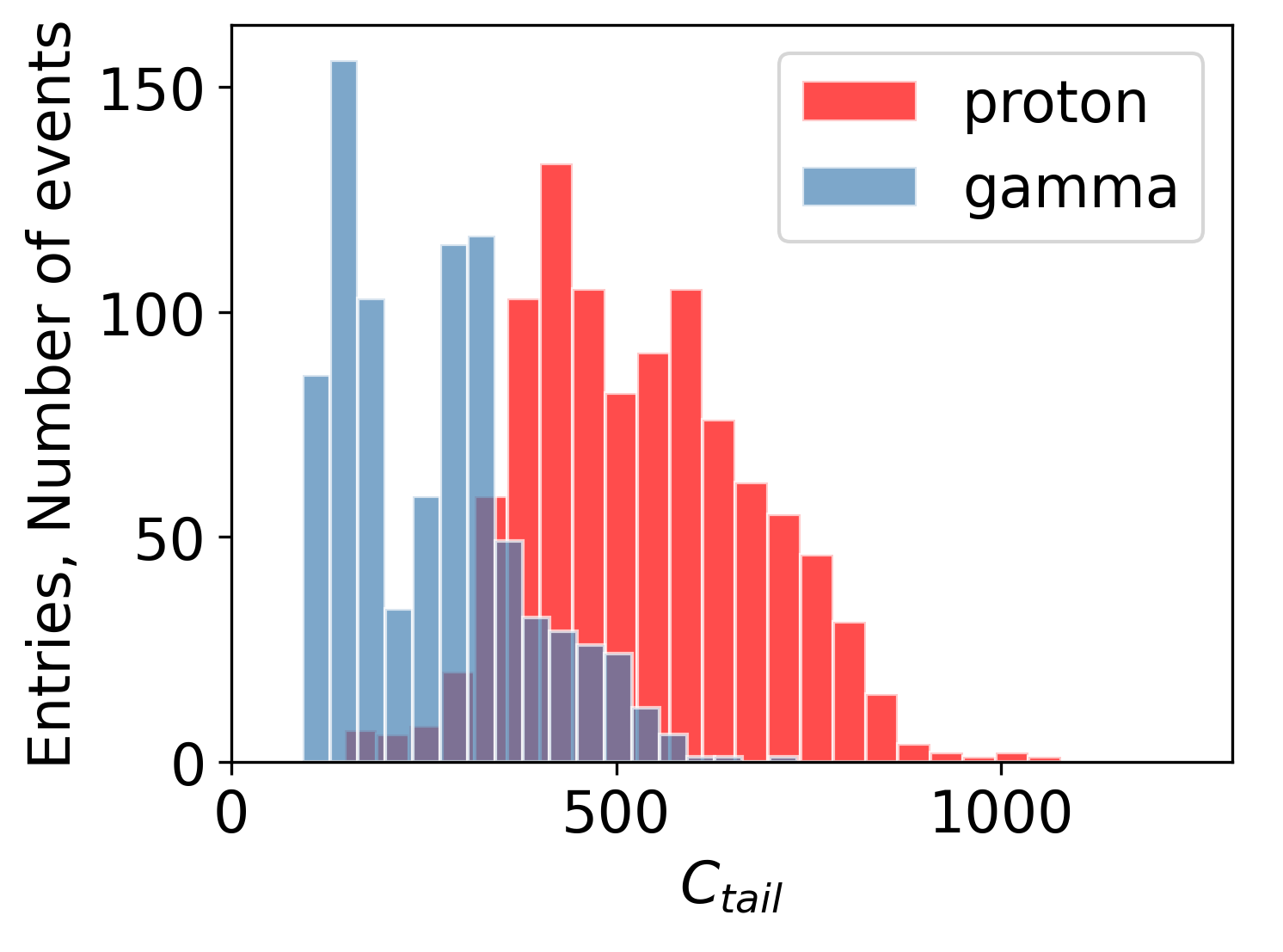}
    \vspace{-0.4cm} 
    \caption{}
  \end{subfigure}
  \begin{subfigure}[b]{0.48\linewidth}
    \centering
    \includegraphics[width=0.85\linewidth]{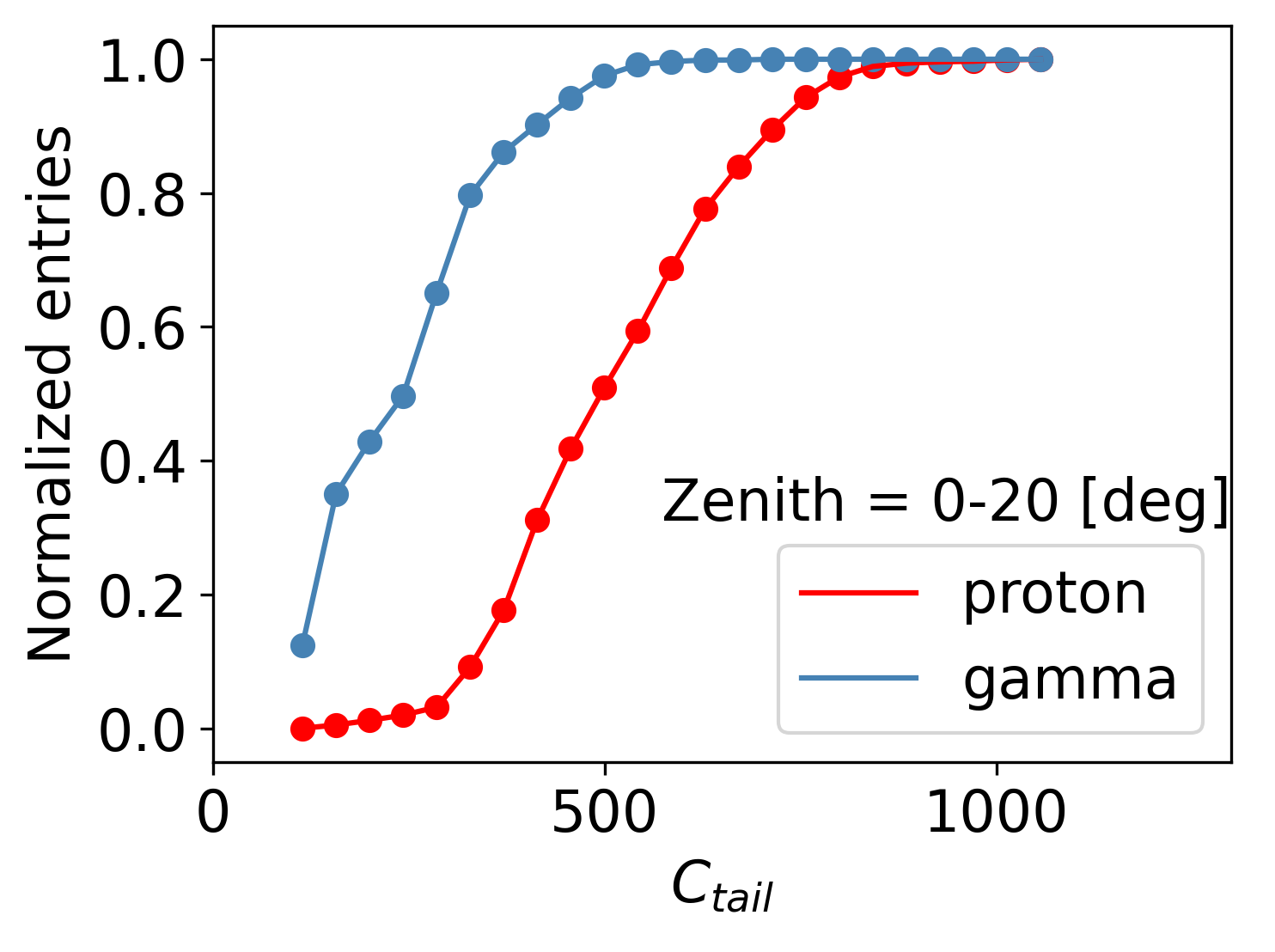}
    \vspace{-0.4cm} 
    \caption{}
  \end{subfigure}
  
  \vspace{-0.3cm} 
  \caption{(a) Histogram of $C_{\text{tail}}$ distribution calculated for events with zenith angles 0-20 [deg], (b) final $C_{\text{tail}}$ cumulative distribution.}
  \label{fig4:-cumulative}
\end{figure}

The final event-level discriminator (Figure \ref{fig4:-cumulative}) is obtained by summing cumulative traces over all active stations in the tail region:

\vspace{-0.9cm} 
\begin{equation}
C_{\text{tail}} = \sum_{i=1}^{M} C_i,    
\end{equation}

\noindent where for each reconstructed event $M$ is the number of detector stations that recorded signals above the detection threshold. The key innovation of the $C_{tail}$ method is the detailed analysis of time-resolved cumulative traces at the individual station level before aggregation. This temporal cumulative analysis captures the time-dependent characteristics of particle shower development in the tail region.


  




\vspace{-0.4cm} 

\vspace{-0.5cm} 

\begin{figure}[H]
  \centering
  \begin{subfigure}[b]{1.0\linewidth}
    \includegraphics[width=1.0\linewidth]{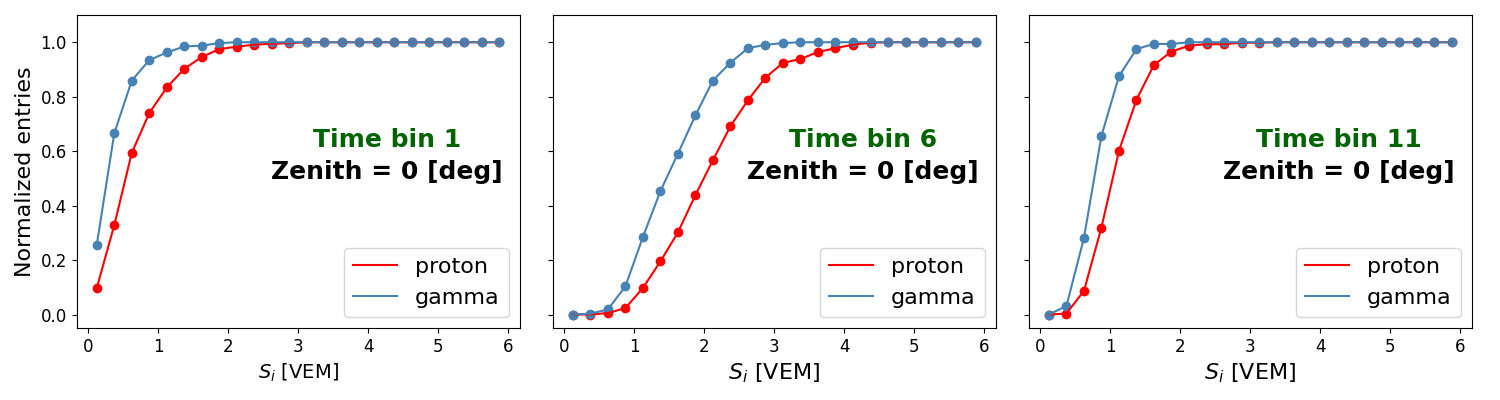}
  \end{subfigure}
  
  \begin{subfigure}[b]{1.0\linewidth}
    \includegraphics[width=1.0\linewidth]{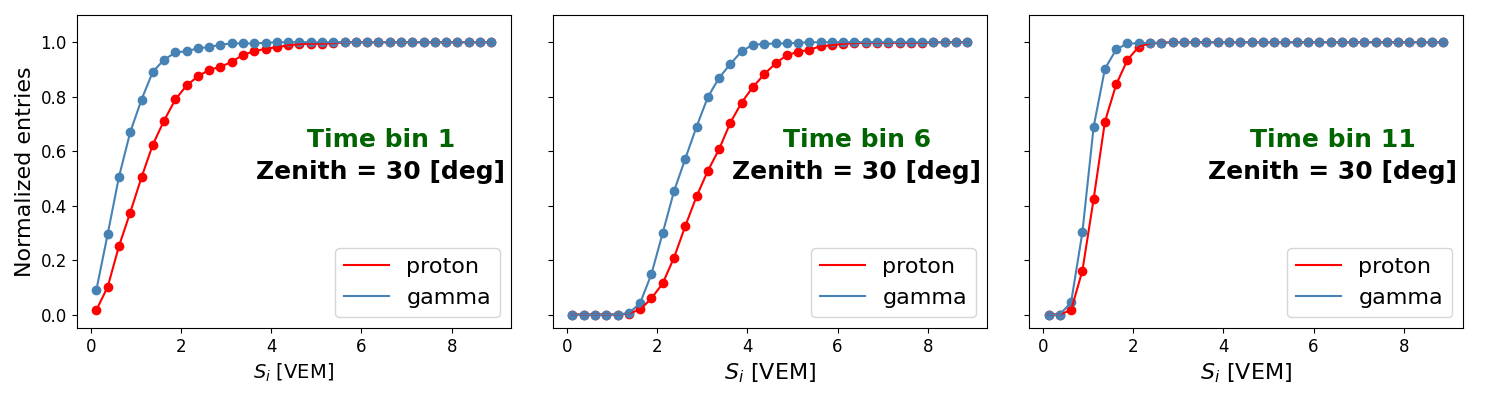}
  \end{subfigure}
    
  \begin{subfigure}[b]{1.0\linewidth}
    \includegraphics[width=1.0\linewidth]{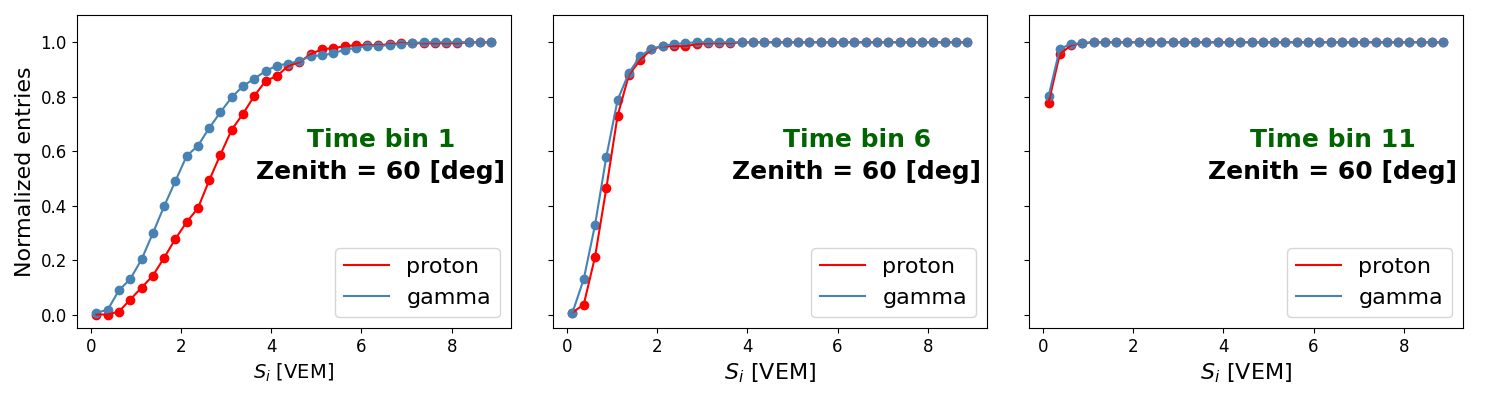}
    
  \end{subfigure}
  
  \vspace{-0.3cm} 
  \caption{Cumulative histograms of instantaneous signal $S_i(k)$ plotted for the same $ring \,3$, but for different time bins, and zenith angles of  0°, 30° and 60°. For showers with a 60° zenith angle, a pronounced difference between gamma-ray and proton showers is observed in the first time bin; however, this distinction diminishes in the subsequent time bins. In contrast, for vertical showers (zenith angle 0°), the difference between these two distributions, indicating discrimination capability, persists up to time bin 11.
  }
  \label{fig5:-cumulative}
\end{figure}

\begin{figure}[H]
    \centering
    \includegraphics[width=1\linewidth]{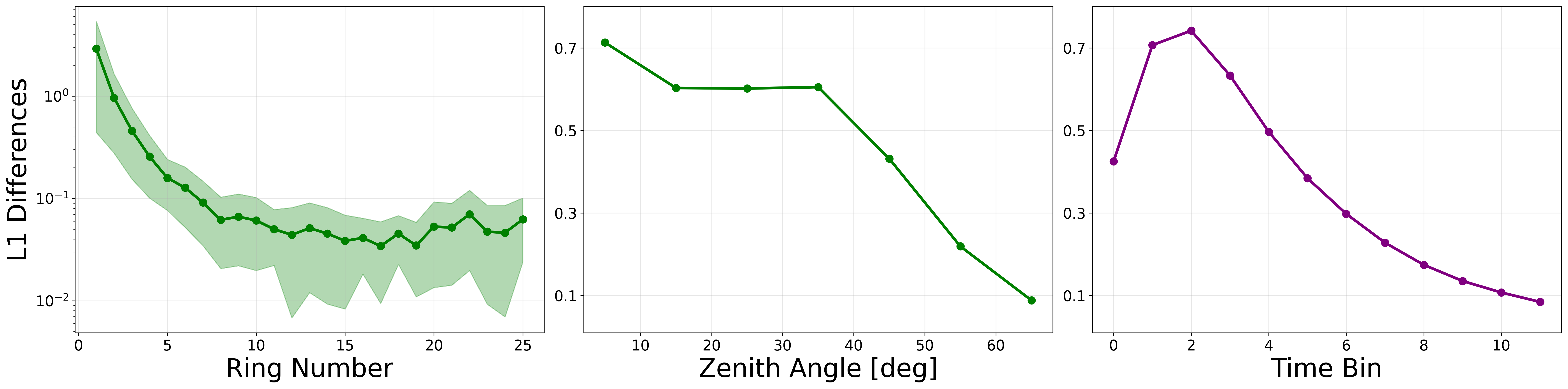}
    \caption{Dependence of the L1 difference between gamma and proton $C_{tail}$ distributions on ring number, zenith angle, and time bin number.}
    \label{fig:6-stats}
\end{figure}

\textbf{Optimization studies.} In the next part, the influence of the selected ranges of time bins, rings, and zenith angles on the $C_{tail}$ discriminant is studied. The discrimination ability of cumulative signal distributions is compared for different time bins within the same ring by analyzing the temporal evolution of cumulative histograms. This temporal cumulative analysis capability represents one of the key advantages of the $C_{tail}$ method. In Figure \ref{fig5:-cumulative}, cumulative distributions are shown for the same ring 3, but for different time bins and zenith angles. For vertical showers (zenith angle 0°), the discrimination capability extends up to time bin 11, while at zenith angle 60°, essential differences between gamma and proton distributions are observed only for the first time bins. As shown in these plots, the difference between the proton and gamma curves decreases as both the zenith angle and time bins increase.
For each zenith angle, there are several cumulative distribution plots with meaningful differences. Thus, the optimal number of time bins that have meaningful differences varies with zenith angle (also can be drawn from Figure \ref{fig:6-stats})).

The discriminating power of different ring-time bin combinations is evaluated to identify optimal configurations for gamma/hadron separation across various zenith angle ranges. To determine the optimal ring, time bin and zenith angle configuration, a statistical estimator (area between curves - the L1 distance \cite{ref10})) is used to maximize the difference between cumulative distributions for gamma and hadron showers (Figure \ref{fig:6-stats}).
The analysis demonstrates that significant differences are present in all rings, which make an important contribution to estimation of $C_{tail}$, but the largest difference is observed in the first 10 rings (200 - 1200 m from shower core). The other plots of Figure \ref{fig:6-stats} are consistent with the previous results, showing that the difference between the proton and gamma curves diminishes when the zenith angle and the number of time bins increase. For zenith angle, the best discriminant range (with maximum difference) is approximately 0°-40°. As for time bin selection, the most promising are the first time bins (1-2), even for 60° zenith angle. The very first time bin 0 is excluded from the optimal values probably because of the noise in this bin reducing the discrimination power. 

 \vspace{-0.3cm} 

\vspace{-0.4cm} 

\begin{figure}[H]
  \begin{subfigure}[b]{0.48\linewidth}
    \centering
    \includegraphics[width=0.75\linewidth]{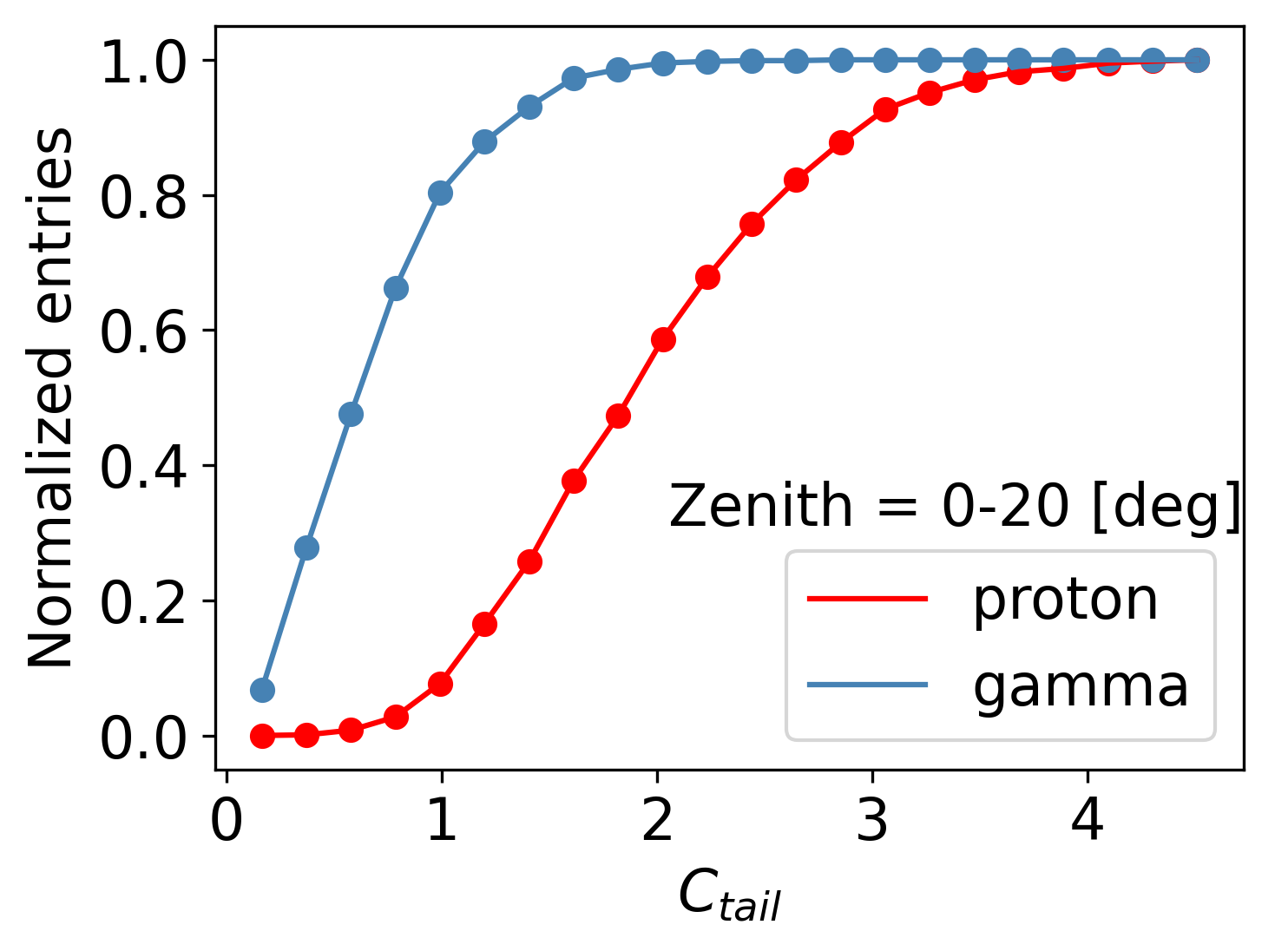}
    \vspace{-0.4cm} 
    \caption{}
  \end{subfigure}
  \begin{subfigure}[b]{0.48\linewidth}
    \centering
    \includegraphics[width=0.75\linewidth]{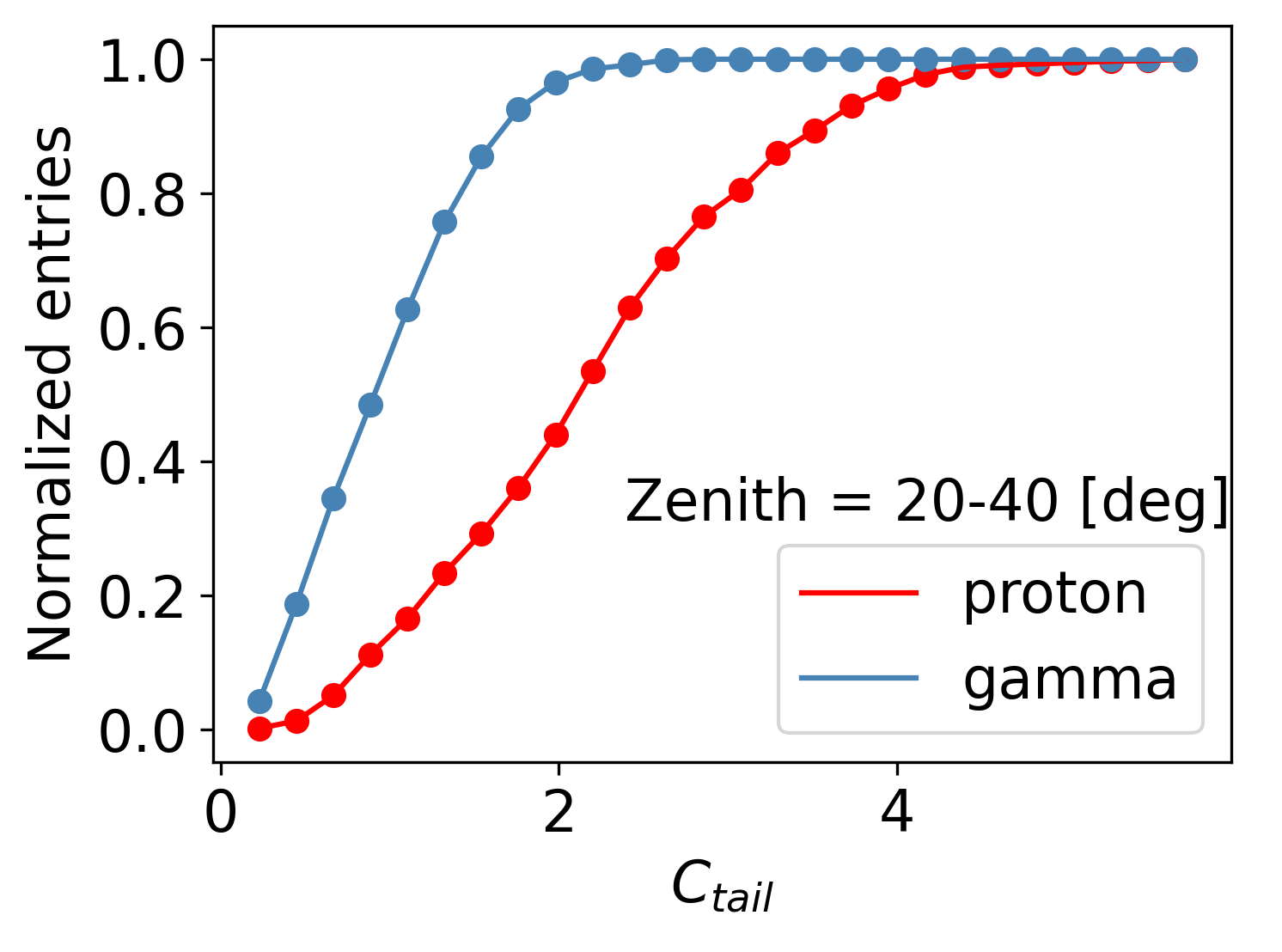}
    \vspace{-0.4cm} 
    \caption{}
  \end{subfigure}

  \vspace{-0.2cm} 
  \caption{(a) and (b) $C_{tail}$ distributions calculated for events with the most promising values of zenith angles 0-40°, time bins 1-2, rings 1-10.}
  \label{fig:7-optimized}
\end{figure}

\vspace{-0.4cm} 
 Based on the most promising estimated values of ring and time bin range, $C_{tail}$ is plotted for the selected rings (1-10), time bins (1-2) and for the most confident zenith angle range (0° - 40°) (Figure \ref{fig:7-optimized}). 
 
\begin{figure}[H]
  \centering
  \begin{subfigure}[b]{1.0\linewidth}
    \includegraphics[width=1.0\linewidth]{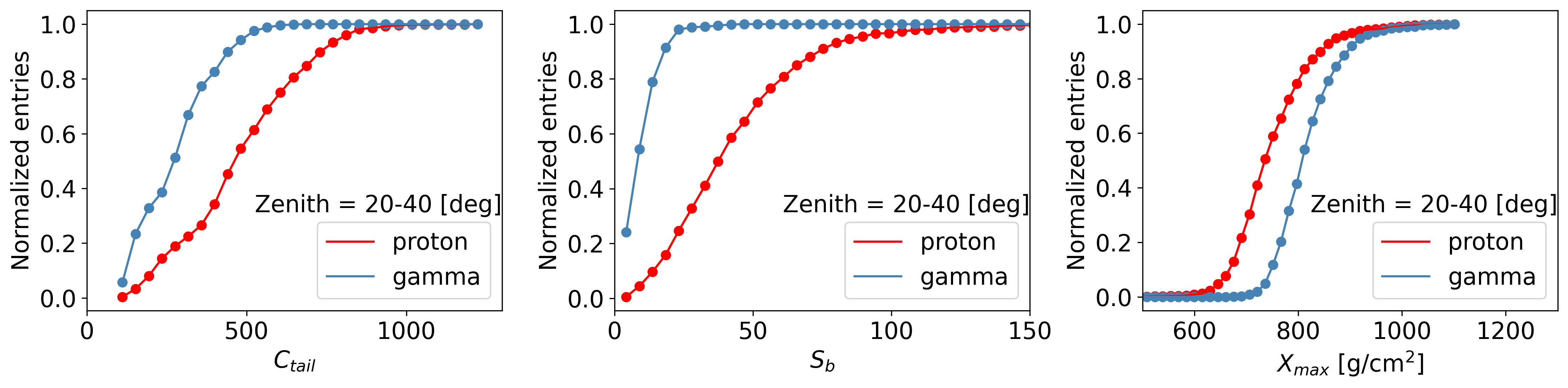}
  \end{subfigure}
    
  \begin{subfigure}[b]{1.0\linewidth}
    \includegraphics[width=1.0\linewidth]{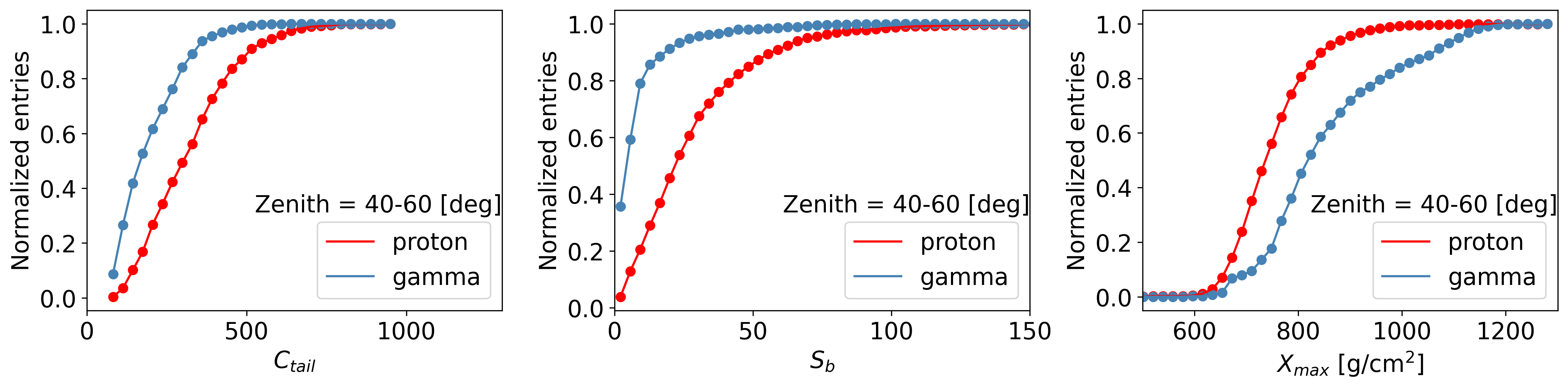}
  \end{subfigure}

  \vspace{-0.3cm} 
  \caption{$C_{tail}$ plotted in comparison to $S_b$ and $X_{max}$ for events of the selected zenith angles.}  \label{fig:8-discriminators}
\end{figure}

\vspace{-0.4cm} 
\textbf{Comparison.} $C_{tail}$ distribution without optimization is shown in Figure \ref{fig4:-cumulative} for zenith angles 0° - 20°. Similarly, the plots in Figure \ref{fig:8-discriminators} are produced for zenith angles 20° - 40° and 40° - 60°. By examining the cumulative behavior of the stations in the tail region, the $C_{tail}$ method demonstrates its effectiveness as a "collective tail analysis" approach. In addition to $C_{tail}$, $S_b$ and $X_{max}$ discriminant variables are plotted for the same events.

\vspace{-0.2cm} 

\vspace{-0.4cm} 
The clear dependence of the discriminants on the zenith angle is observed in these figures. All these variables $C_{tail}$, $S_b$ and $X_{\max}$ show essential discriminant power and are promising for various gamma/hadron discriminant study analysis. They can also be used as complementary variables in the investigation of the best discriminant variable, as they all are of a different nature.

\section{Summary}

In this paper, a trace-based discriminant variable $C_{tail}$ is presented and evaluated using a detector array with 750 m spacing between detectors. The $C_{tail}$ discriminant variable demonstrates strong separation capabilities between gamma and proton events, and can be used alongside the conventional variables $S_b$ and $X_{\max}$.
The $C_{tail}$ method represents a novel approach to gamma/hadron discrimination that focuses on the same physical phenomenon as $P_{tail}$ (tail region anomalies) but uses cumulative signal analysis instead of probability-based analysis. $C_{tail}$ uses multi-level cumulative analysis to characterize the temporal evolution of signal traces at individual stations before aggregating them into event-level discriminants. In this way it operates at the shower level rather than individual station level.
$C_{tail}$ provides complementary information to existing methods, shows optimal performance in the 200-1200m tail region for zenith angles 0°-40° and time bins 1-2.

$C_{tail}$ as well as $S_b$ and $X_{\max}$ variables are valuable tools for cosmic ray air-shower experiments focused on primary particle identification. These discriminant variables will be particularly valuable in advanced analysis techniques, for searching for flares of ultra-high-energy cosmic rays, such as direction-time clustering analysis described in \cite{ref11}. The trace-based methodology presented provides an effective strategy for enhancing primary particle identification precision in extensive air shower arrays, thereby advancing our comprehension of the most energetic astrophysical processes in the universe.

\section{Acknowledgments}
We want to acknowledge the support in Poland from the National Science Centre, grants No. 2020/39/B/ST9 /01398 and 2022/45/B/ST9/02163 as well as from the Ministry of Science and Higher Education, grant No. 2022/WK/12.

\begin{thebibliography}{99}
\bibitem{ref1} Billoir, P., Pierre Auger Collaboration, "Photon/hadron Discrimination of the Auger Observatory", ICRC 2001, Hamburg.
\bibitem{ref2} Ros, G. et al., "A new composition-sensitive parameter for ultra-high energy cosmic rays", Astropart. Phys. 35 (2011) 140.
\bibitem{ref3} Conceição. R. et al., "High resolution gamma/hadron and composition discriminant variable for water-Cherenkov detector cosmic-ray observatories", Phys. Rev. D 110 (2024) 023033.
\bibitem{ref4} Costa, P., "$Ptail^T$: a new gamma-hadron discriminator for UHE searches", Auger Malargüe Collaboration Meeting, Nov. 2024.
\bibitem{ref5} Heck, D., et al., "CORSIKA: A Monte Carlo code to simulate extensive air showers", FZKA 6019 (1998).
\bibitem{ref6} Mariş, I. C., Pierre Auger Collaboration, "The AMIGA infill detector of the Pierre Auger Observatory: performance and first data", ICRC 2011, Beijing.
\bibitem{ref7} Pierog, T. and Werner, K., "EPOS LHC: Test of collective hadronization with data measured at the CERN Large Hadron Collider", PoS(ICRC2023)230.
\bibitem{ref8} Bleicher, M., et al., "Relativistic hadron-hadron collisions in the ultra-relativistic quantum molecular dynamics model", J. Phys. G 25 (1999) 1859.
\bibitem{ref9} Argir{\`o}, S., et al., "The Offline Software Framework of the Pierre Auger Observatory", Nucl. Instrum. Methods A 580 (2007) 1485.
\bibitem{ref10} Wasserman, L., "All of Nonparametrics: A Concise Course in Nonparametric Statistical Inference", Springer (2006).
\bibitem{ref11} Stasielak, J., et al., "Advanced techniques of searching for flares of ultra-high-energy photons", PoS(ICRC2025)400, to be published.
\end{thebibliography}
\end{document}